\documentclass[11pt]{article}
\usepackage{amsfonts,amsmath,amssymb}
\usepackage{graphicx}

\newfont{\twelvemsb}{msbm10 scaled\magstep1}
\newfont{\eightmsb}{msbm8}
\newfam\msbfam
\textfont\msbfam=\twelvemsb \scriptfont\msbfam=\eightmsb
\catcode`\@=11
\def\Bbb{\ifmmode\let\next\Bbb@\else
\def\next{\errmessage{Use \string\Bbb\space only in math mode}}\fi\next}
\def\Bbb@#1{{\fam\msbfam{{#1}}}}

\newcommand{\be}{\begin{equation}}
\newcommand{\ee}{\end{equation}}
\newcommand{\ba}{\begin{eqnarray}}
\newcommand{\ea}{\end{eqnarray}}

\begin{document}
\sloppy
\renewcommand{\thefootnote}{\fnsymbol{footnote}}
\newpage
\setcounter{page}{1} \vspace{0.7cm}

\vspace*{1cm}
\begin{center}
{\bf From Painlev\'e equations to ${\cal N}=2$ susy gauge theories: prolegomena}\\
\vspace{1.8cm} {\large Davide Fioravanti $^a$ and Marco Rossi $^b$
\footnote{E-mail: fioravanti@bo.infn.it, rossi@cs.infn.it}}\\
\vspace{.5cm} $^a${\em Sezione INFN di Bologna, Dipartimento di Fisica e Astronomia,
Universit\`a di Bologna\\
Via Irnerio 46, 40126 Bologna, Italy} \\
\vspace{.3cm} $^b${\em Dipartimento di Fisica dell'Universit\`a della Calabria and
INFN, Gruppo collegato di Cosenza\\
Arcavacata di Rende, 87036 Cosenza, Italy} \\
\end{center}
\renewcommand{\thefootnote}{\arabic{footnote}}
\setcounter{footnote}{0}
\begin{abstract}
We study the linear problems in $z,t$ (time) associated to the Painlev\'e III$_3$ and III$_1$ equations when the Painlev\'e solution $q(t)$ approaches a pole or a zero. In this limit the problem in $z$ for the Painlev\'e III$_3$ reduces to the modified Mathieu equation, while that for the Painlev\'e III$_1$ to the Doubly Confluent Heun Equation. These equations appear as Nekrasov-Shatashvili quantisations/deformations of Seiberg-Witten differentials for $SU(2)$ ${\cal N}=2$ super Yang-Mills gauge theory with number of flavours $N_f=0$ and $N_f=2$, respectively. These results allow us to conjecture that this link holds for any Painlev\'e equation relating each of them to a different matter theory, which is actually the same as in the well-established Painlev\'e gauge correspondence, but {\it with another deformation ($\Omega$-background)}. An explicit expression for the dual gauge period (and then prepotential) is also found. As a by-product, a new solution to the connexion problem is illustrated.

\end{abstract}
\vspace{1cm} 
{\noindent {\it Keywords}}: ODE/IM correspondence;  Painlev\'e equations; SYM theories
\newpage

\section{Introduction}
\label{intro}
\setcounter{equation}{0}

The original Ordinary Differential Equation/Integrable Model (ODE/IM) correspondence \cite {ODE-IM1} finds useful and effective relations among connection and monodromy coefficients (of the ODE with monomial potential plus possibly angular momentum) and proves that they equal, respectively, the $Q$ and $T$ functions of a suitable quantum integrable model. Actually, the latter was originally a 2D Conformal Field Theory (CFT) minimal model stemming from one ODE, whilst this derivation was extended by \cite{LZ} from a (classical) sinh-Gordon (ShG) Lax pair to quantum sine-Gordon (SG) theory and further generalised to theories with many massive moduli by \cite{FRS} with a description of all the integrability structures. In the other direction, from the classical to the quantum theory, the derivation was reversed and extended to the case with many masses in \cite{FR1}, by going from the basic equation of an integrable quantum theory, the $QQ$-system, to the formulation of the corresponding classical Lax problem and then giving a way to set in use classical physics and a reason for the correspondence to hold. Actually, as analysed in \cite{FRS} in some generality, the $QQ$-system is equivalent to the $TQ$-system, provided the $T$ function is properly defined.   

Schematically, we may think of the full theory, in the classical to quantum direction, as made up of two parts: the first one consists in the derivation of the quantum $QQ$-system or the $TQ$-system, the basis of the IM, from the classical Lax pair: we name this part 'Parent-ODE/IM' (P-ODE/IM) correspondence. Moreover, at this point a second part emerges as very intriguing to build up: since the $TQ$-system is a finite difference equation, we would like to set up a systematic study of its connection and monodromy coefficients: quite intuitively we will name this part 'Child-ODE/IM' (C-ODE/IM) correspondence. This idea was also motivated by the study \cite{FR1} which essentially converts the $TQ$-system to the Lax pair.

In this letter we start a study of this in a peculiar limit case, very significant in itself. As P-ODE/IM we consider the Lax pair corresponding to quantum SG theory (only one mass) on a cylinder with circumference $R$ in the vacuum \cite {LZ}. The Lax pair contains derivatives with respect to variables $w, \bar w$ of a configuration space, a spectral parameter\footnote {Here we adopt different notations with respect to \cite {FR1}: we call $y$ the spectral parameter of the ODE/IM correspondence and ${\cal Q}_\pm (y), {\cal T}(y)$ the related $Q$ and $T$-functions: these in \cite {FR1} are called $Q_\pm (\theta =y/2),T (\theta =y/2)$.} $y$ and a 'potential' specified by one solution $\hat \eta (w,\bar w)$ (as for the quantum ground state) of the classical ShG equation (arising from compatibility of the Lax pair). Crucially, from the limit of the wave functions $\psi _\pm$ to a {\it special point} $w=w_0=-\frac {r}{4\cos \frac {\pi \beta ^2}{2(1-\beta ^2)}}$, with $\beta ^2$ the coupling of SG and $r=MR$, with $M$ the soliton mass, one obtains ${\cal Q}_\pm(y)$, the $Q$-functions of the ground state of SG (with also masses in \cite {FR1}). 
They satisfy 
quasiperiodicity
$ {\cal Q}_{\pm}(y +2i\tau)=e^ {\pm i\pi \left (l+\frac  {1} {2} \right )}{\cal Q}_ {\pm}(y)$, with
$\tau=\pi/(1-\beta ^2)$ and $-\frac {1}{2}<l<\frac {1}{2}$ and 
difference equations in $y$, belonging to the P-ODE/IM part, the $TQ$- and $QQ$-systems. 
At this point the C-ODE/IM analysis would be very interesting, namely finding exact relations among the monodromy and connection coefficients of these functional relations. In Section \ref {app} we are successful in the special limit in which the quantum SG coupling $\beta ^2 \rightarrow 0$, with $r$ and $l$ fixed \footnote {This differs, importantly, from the classical limit.} as the $TQ$-system reduces to the modified Mathieu equation (MME) for ${\cal Q}_\pm(y)$. Moreover the corresponding ShG Lax pair (and then ShG classical equation) goes to its radial restriction, i.e. the Painlev\'e III$_3$ one. 
In the Painlev\'e theory the point $w=w_0$ occurs at time $t=r$ of the closest Painlev\'e movable pole and in fact the first Lax linear problem becomes the MME at $t=r$. In Section \ref {app} we revisit this proof.  

\vspace{1cm}

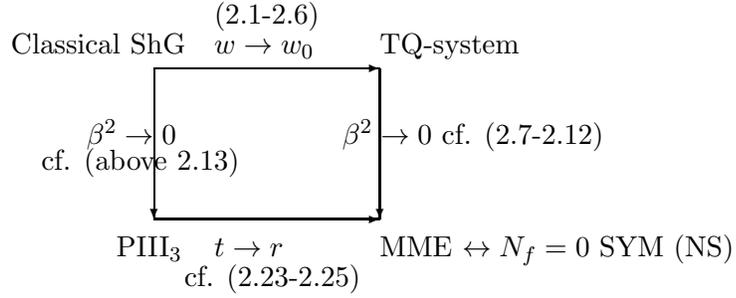
\begin{center}
\setlength{\unitlength}{1cm}
\begin{picture}(3,2)(0,0)
\put(-0.5,-0.5){PIII$_3$}
\put(-0.9,1){$\beta ^2 \rightarrow 0$}
\put(-1.5,0.65){cf. (above 2.13)}
\put(2.5,1){$\beta ^2 \rightarrow 0$ cf. (2.7-2.12)}
\put(0,0){\vector(3,0){3}}
\put(0,2){\vector(1,0){3}}
\put(0,2){\vector(0,-1){2}}
\put(3,2){\vector(0,-1){2}}
\put(3,2.2){TQ-system}
\put(0.8,2.6){(2.1-2.6)}
\put(0.8,2.2){$w\rightarrow w_0$}
\put(-1.9,2.2){Classical ShG}
\put(3,-0.5){MME $\leftrightarrow N_f=0$ SYM (NS)}
\put(0.8,-0.5){$t\rightarrow r$}
\put(0.4,-0.9){cf. (2.23-2.25)}

\put(-6.4,-1.5){P-ODE/TQ-system (IM) correspondence: general case (upper line), Painlev\'e limit (lower line)}
\end{picture}
\end{center}

\vspace {1.5 cm}

Interestingly, it is known \cite {AY} that the MME realises the quantisation of Seiberg-Witten (SW)
\cite {SW} differential for pure (number of flavours $N_f=0$) $SU(2)$ ${\cal N}=2$ super Yang-Mills (SYM) gauge theory in the Nekrasov-Shatashvili (NS) background \cite {NS}.
In the last part of this letter the relation between Painlev\'e equations and ${\cal N}=2$ SYM theories in the NS background is extended to cases with flavours. 
In particular (Section \ref {P31}) we consider the Painlev\'e III$_1$ equation and discuss its linear problem near a simple pole (or simple zero) of its solution. In an analogous fashion as in the Painlev\'e III$_3$ case, we find that one of the Lax operators reduces to the Doubly Confluent Heun Equation (DCHE), i.e. the ODE describing ${\cal N}=2$ SYM with $N_f=2$ and conjecture that this holds for any Painlev\'e equation to reproduce a mother theory with different flavour multiplets $N_f=0,1,2,3,4$. 
In other words, what we prove and propose is an interesting novel Painlev\'e/gauge correspondence (lower line of the figure above).
It is of particular importance as the Painlev\'e flow is {\it isomonodromic} and then studying the monodromies of the quantum gauge differential (in the NS background) is the same as the Painlev\'e's ones. On the other hand, a known correspondence \cite {BLMST} links Painlev\'e theory - i.e. a limit case of the P-ODE/IM theory - to the gauge theory in a different background, the self-dual (SD) one: this point deserves further study in the future. 

\section{ODE/IM $TQ$-system and Painlev\'e III$_3$/NS $N_f=0$ correspondence}
\label{app}
\setcounter{equation}{0}

We move from SG model on a cylinder ($\varphi (x+R,t)=\varphi (x,t)$) and with Lagrangian 
$\mathcal{L}=\frac  {1}{16\pi} \left [ (\partial _t \varphi)^2-(\partial _x \varphi )^2 \right ] +2\mu \cos \beta \varphi  \, .
$
The P-ODE/IM relates the quantum SG model in the vacuum to the Lax linear problems
\be
D(w,\bar w)\Psi _{\beta ^2}(w,\bar w) =0 \, ,\,\,\,  \bar D (w,\bar w) \Psi _{\beta ^2}(w,\bar w)=0 \, , \label{Laxpr}
\ee
with $D(w,\bar w)$ and $\bar D(w,\bar w)$ given by
\be
D(w,\bar w)=\frac{\partial}{\partial w}+\frac{1}{2}\frac{\partial \hat \eta }{\partial w}\, \sigma ^3 -e^{\frac {y}{2} +\hat \eta  } \sigma ^+ - e^{\frac {y}{2} -\hat \eta  } \sigma ^- \, , \quad \bar D(w,\bar w)=\frac{\partial}{\partial \bar w}-\frac{1}{2}\frac{\partial \hat \eta  }{\partial {\bar w}}\, \sigma ^3 -e^{-\frac {y}{2} +\hat \eta  } \sigma ^- - e^{-\frac {y}{2} -\hat \eta  } \sigma ^+ \, , \label{conn}
\ee
with $\hat \eta (w,\bar w)$ a 2D classical scalar field satisfying the classical ShG equation
\be
\frac{\partial ^2 \hat \eta  }{\partial w \partial {\bar w}} = 2\sinh 2\hat \eta \, .
\label{etashg}
\ee
As shown in \cite {FR1}, the regularised choice
\be
 \Psi _{\beta ^2}(w,\bar w)=\begin{pmatrix}  e^{\frac {\hat \eta}{2}}\tilde \psi _+ (w,\bar w)\\ - e^{-\frac {\hat \eta}{2}}\tilde \psi _- (w,\bar w) \end{pmatrix}
\label  {Psitildepsi} \, , 
\ee
where $\tilde \psi _\pm $ (the wave functions used in \cite {FR1}, where they are called $\psi _\pm$) are particular Jost functions, reproduces, once computed at the logarithmic singularity $w=w_0$ for $\hat \eta$ ($\tilde \psi _\pm$ are diverging, but they are compensated by $e^{\pm \hat \eta /2}$, respectively), the $Q$-functions:
\be
\Psi _{\beta ^2}(w_0,w_0)= e^{-i\pi \frac{\sigma ^3}{4}}
\frac{1}{\sqrt{\cos \pi l}}
\left ( \begin{array}{c}
e^{-\frac {y}{2} l}{\cal Q}_+\left(y+i\tau\right ) \\
e^{\frac {y}{2} l}{\cal Q}_-\left(y+i\tau \right )
\end{array} \right ) \, . \label{hatXi-Q}
\ee
The most important relation between connexion and monodromy coefficients is the Baxter's $TQ$-system
\be
{\cal T}(y ){\cal Q}_\pm (y)={\cal Q}_\pm (y+2i\pi \xi)+{\cal Q}_\pm(y-2i\pi \xi) \, , \quad \xi =\beta ^2/(1-\beta ^2) \, .
\label  {TQmother}
\ee
We are going to show that the MME arises in the Parent theory (the SG one) as limit of the $TQ$-relation, when the quantum SG coupling $\beta ^2 \rightarrow 0$, with $r,l$ fixed. 
As $\xi \rightarrow 0$, $T$ and $Q$ functions expand as
\ba
{\cal T}(y)&=&t^{(0)}(y)+\xi t^{(1)}(y)+\xi ^2 t^{(2)}(y)+O(\xi ^3) \, , \\
{\cal Q}_\pm(y)&=&q^{(0)}_\pm(y)+\xi q^{(1)}_\pm(y)+\xi ^2 q^{(2)}_\pm(y)+O(\xi ^3) \, .
\ea
At order zero (\ref {TQmother}) implies
$ t^{(0)}(y)q^{(0)}_\pm(y)=2 q^{(0)}_\pm(y)$, 
which (supposing $q^{(0)}_\pm(y)\not= 0$) gives $t^{(0)}(y)=2$.
At order $\xi$ 
$2q^{(1)}_\pm(y)=t^{(1)}(y)q^{(0)}_\pm(y)+ 2 q^{(1)}_\pm(y)$, 
which gives $t^{(1)}(y)=0$.
At order $\xi ^2$ it implies
\be
4\frac {d^2}{d y^2}q^{(0)}_\pm(y)=-\frac {t^{(2)}(y) }{\pi ^2} q^{(0)}_\pm(y) \, , 
\label {schroe}
\ee
which is a Schr\"{o}dinger-like equation for $q^{(0)}_\pm(y)$ with 'potential' $t^{(2)}(y) $.
The function $t^{(2)}(y) $ is $2\pi i$ periodic, 
$t^{(2)}(y+2i\pi)=t^{(2)}(y)$, as a consequence of the quasi-periodicity of ${\cal Q}_\pm$ and the $TQ$-relation (\ref {TQmother}) and has parity
$t^{(2)}(y)=t^{(2)}(-y)$: then, in the general 
the functional form $t^{(2)}(y)=\sum\limits _{n=0}^{+\infty} A_n \cosh ny$. The constants $A_n$ are forced by the limits Re$y \rightarrow \pm \infty$ from those of ${\cal Q}_\pm (y)$, via the $TQ$-system. For the asymptotic expansions hold 
\be
\ln {\cal Q}_\pm\left (y+{i\pi (1+\xi)} \right ) \simeq \frac {re^{\frac {\epsilon y}{2}}}{4\cos \frac {\pi \xi}{2}}
\pm A
 -I_1 ^\pm e^{-\frac {\epsilon y}{2}}+o\left (e^{-\frac {\epsilon y}{2}} \right ) \, , 
\ee
as $\textrm{Re}y\rightarrow +\epsilon \infty$, $\epsilon =\pm 1$, within $|\textrm{Im}y|<\pi (1+\xi)$, where $A$ is a constant not relevant for our aims and $I_1^\pm $ are vacuum eigenvalues of the first conserved local charges of SG. Actually, what enters the $TQ$-system in the small $\xi$ limit is the sum $I_1^++I_1^-=-\frac {r}{2}-\frac {2\pi}{\xi r} {\cal F}$, with ${\cal F}$ the free energy of the SG model:
\be
t^{(2)}(y)=\frac {r^2\pi ^2 }{8}\cosh y-\frac {r\pi ^2}{4} \left (-\frac {r}{2}-\frac {2\pi}{\xi r} {\cal F} \right )  \, , \quad 0<\textrm {Im} y<\pi \, .
\ee
Eventually (\ref {schroe}) becomes the MME
\be 
\frac {d^2}{dy^2}q^{(0)}_\pm (y)=\left ( -\frac {r^{2}} {32} \cosh y +P^2 \right ) q^{(0)}_\pm (y)
\label {math} \, ,
\ee 
with the constant
$P^2=\frac {r}{16}(I_1^++I_1^-) = -\frac {r}{16} \left (\frac {r}{2}+\frac {2\pi}{\xi r} {\cal F} \right ) $
containing the quantum SG free energy (for $\xi \rightarrow 0$). 
Quasi-periodicity of $ {\cal Q}_\pm(y)$ entails that of $q^{(0)}_\pm (y+2i\pi)=e^ {\pm i\pi \left (l+\frac  {1} {2} \right )}q^{(0)}_\pm (y)$, which identifies them as Floquet solutions of the MME, $q^{(0)}_\pm (y+2i\pi)=e^ {\pm 2\pi i k}q^{(0)}_\pm (y)$, with Floquet index 
$k=\left (\frac {l}{2} +\frac {1}{4}\right ) $.

\medskip

To study the results of the same limit on the classical theory, it is convenient to 
use polar coordinates $t,\phi$, defined around the singularity $w_0$: namely $w-w_0=\frac {t}{4}e^{i\phi}$, $\bar w-\bar w_0=\frac {t}{4} e^{-i\phi}$, since 
in this limit the configuration space of $w$ reduces to the segment $-r/4=w_0<w<0$ (together with the half-plane Re$w>0$). As a consequence, $\hat \eta$ depends only on $t=4|w-w_0|$, while the ShG equation (\ref {etashg}) computed at $w-w_0$ collapses into the Painlev\'e III$_3$ equation, $\frac {d^2\hat \eta}{dt^2}+\frac {1}{t}\frac {d\hat \eta}{dt}=\frac {1}{2}\sinh 2\hat \eta $. We then define wave functions, proper of the Painlev\'e limit, as
\be
\Psi (t,\phi)=\begin{pmatrix}  e^{\frac {\hat \eta}{2}}\tilde \psi _+ (t,\phi)\\ - e^{-\frac {\hat \eta}{2}}\tilde \psi _- (t,\phi) \end{pmatrix}
=\lim _{\beta ^2 \rightarrow 0} \Psi _{\beta ^2}(w-w_0,\bar w-w_0) \, ,
\ee
so that, with the computation of (\ref {Laxpr}) upon replacing $w$ with $w-w_0$, they satisfy 
\ba
\lim _{\beta ^2 \rightarrow 0} [(w-w_0)D(w-w_0)  -(\bar w -w_0) \bar D (\bar w-w_0)]  \Psi _{\beta ^2} (w-w_0,\bar w-w_0) = (-i \partial _{\phi}-2{\bf A}_y)\Psi (t,\phi)  = 0
\, , \label  {lprbtheta0} \\ 
\lim _{\beta ^2 \rightarrow 0} [(w-w_0) D(w-w_0)+(\bar w -w_0) \bar D(\bar w-w_0)]  \Psi  _{\beta ^2} (w-w_0,\bar w-w_0)=t  (\partial _t - {\bf A}_t)\Psi  (t,\phi) = 0 \, , \label {lprbti0}
\ea
with
\be 
{\bf A}_t=\begin{pmatrix} 0 & \frac {1}{2}\cosh (\frac {y}{2}+i\phi+\hat \eta) \\  \frac {1}{2}\cosh (\frac {y}{2}+i\phi-\hat \eta)  & 0 \end{pmatrix} \, , \quad 
{\bf A}_y= \begin{pmatrix} -\frac {t}{4}\frac {d\hat \eta}{dt}  & \frac {t}{4}\sinh (\frac {y}{2}+i\phi+\hat \eta) \\  \frac {t}{4}\sinh (\frac {y}{2}+i\phi-\hat \eta)  & \frac {t}{4}\frac {d\hat \eta}{dt} \end{pmatrix} \label {Aours} \, . 
\ee
with $\hat \eta (w\rightarrow w-w_0)=\hat \eta (t)$.
Since the dependence of ${\bf A}_t, {\bf A}_y$ on $y,\phi$ is through the combination $y+2i\phi$, we can consider solutions $\Psi $ of both two linear problems (\ref  {lprbtheta0}, \ref {lprbti0}) depending
only on $t$ and $y+2i\phi$, which then become 
$ (\partial _{y}-{\bf A}_y)\Psi (t,y+2i\phi)=  0 $, 
$(\partial _t - {\bf A}_t)\Psi  (t,y+2i\phi) = 0$, namely the Lax pair of the Painlev\'e III$_3$ equation as isomonodromic flow 
\cite {FR1}. In fact, 
the change of coordinates $s=\left (\frac {t}{8} \right )^4$, $z=\frac {t^2}{64}e^{y+2i\phi}$ and the redefinitions 
\be
q=\frac {1}{2^6}e^{-2\hat \eta}t^2 \, , \quad \psi (s,z)= \footnotesize {\begin{pmatrix} 0 & \sqrt {\frac {t}{8}} e^{\frac {y+2i\phi+2\hat \eta }{4}} \\
\sqrt {\frac {8}{t}} e^{-\frac {y+2i\phi+2\hat \eta }{4}} & 0 \end{pmatrix} } \Psi (t,y+2i\phi) = \begin{pmatrix} - \sqrt {\frac {t}{8}} e^{\frac {y+2i\phi}{4}}\tilde \psi _-(t,y+2i\phi) \\  \sqrt {\frac {8}{t}}e^{-\frac {y+2i\phi}{4}}\tilde \psi _+(t,y+2i\phi)  \end{pmatrix}\equiv  \begin{pmatrix} \psi _-(s,z) \\ \psi _+(s,z)  \end{pmatrix} \, , \label {redef}
\ee
put them in a more traditional form (c.f. for instance \cite {BLMST}): 
\ba
&& (\partial _z -{\bf A}_z )\psi (s,z) =0 \, , \quad (\partial _s -{\bf A}_s) \psi (s,z) =0 \label {tradlax}
\\
&& {\bf A}_z=\begin{pmatrix} \frac {pq}{z} & 1-\frac {s}{zq} \\ \frac {1}{z}-\frac {q}{z^2} & -\frac {pq}{z} \end{pmatrix} \, , \quad 
{\bf A}_s=\begin{pmatrix}  0 & \frac {1}{q} \\ \frac {q}{sz} & 0 \end{pmatrix}
\label {bfAs} \, , \quad pq=\frac {1}{2}-\frac {s\dot {q}}{2q} \, , \quad \dot q=\frac {dq}{ds} \, .
\ea
In terms of $q(s)$  Painlev\'e III$_3$ equation reads
$\ddot {q}=\frac {\dot {q}^2}{q}-\frac {\dot {q}}{s}+\frac {2q^2}{s^2}-\frac {2}{s}$.
This has the symmetry $q \rightarrow s/q$, which means $\hat \eta \rightarrow -\hat \eta$. With the (a little abusive) identification,
$\psi _\pm (s,z)=\psi _\pm (t,y+2i\phi)$ enter the vector solution 
\be
 \Psi (t,y+2i\phi)=\begin{pmatrix}  e^{\frac {\hat \eta}{2}+\frac {y+2i\phi}{4}}\sqrt {\frac {t}{8}}\psi _+(t,y+2i\phi)\\
e^{-\frac {\hat \eta}{2}-\frac {y+2i\phi}{4}}\sqrt {\frac {8}{t}}\psi _-(t,y+2i\phi) \end{pmatrix}
\label  {Psi-psi} \, .
\ee
Furthermore, the off-critical ODE/IM teaches us in general how to obtain the $Q$-functions as limit to the ShG solution singularity $w=w_0$ (\ref {hatXi-Q}), which importantly becomes here, at $\beta ^2 \rightarrow 0$ (supposing the commutation of the two limits), the limit $w-w_0=(t/4) e^{i\phi} \rightarrow w_0=-r/4$  \cite {FR1}: 
\be
 \lim \limits _{\substack {t\rightarrow r \\ \phi \rightarrow \pi}}  \Psi \left (t, y+2i\phi \right)= 
i\left ( \begin{array}{c}
e^{\frac{y}{4}+\hat \eta}q^{(0)}_+\left(y+i\pi\right ) \\
e^{-\frac{y}{4}}q^{(0)}_-\left(y+i\pi\right )
\end{array} \right ) \, , \label{hatXi-Q2}
\ee
with $q_\pm^{(0)}$ satisfying the MME (\ref {math}). 
This result  proves immediately its usefulness by suggesting a way\footnote{Similar calculations (for the Painlev\'e equations I and IV) were obtained in \cite {MAS} with different aims.}, generalisable to any Painlev\'e, in which the MME has to arise in the context of the Lax problems (\ref {tradlax}, \ref {bfAs}): as limit of the first Lax problem (\ref {bfAs}) to a special time $t=r$ of the isomonodromic flow, namely a pole or zero of a Painlev\'e solution. In order to discuss this, we use the time 
$t$ and start from the equivalent equation ($q'=\frac {dq}{dt}$)
\be
q''+\frac {q'}{t}=\frac {{q'^2}}{q}+\frac {32q^2}{t^2}-\frac {t^2}{128} \, . \label {paintau}
\ee
It is well known that a zero or a pole  (related by the symmetry $q \rightarrow s/q$) of a solution $q(t)$ to (\ref {paintau}) in $t=r$ is double. 
Therefore, without losing generality, we consider a double zero, insert the generic form $q(t)=\gamma (t-r)^2+\rho (t-r)^3+\kappa (t-r)^4 +O\left ( (t-r)^5 \right )$ in (\ref {paintau}) and find $\gamma$ and $\rho$ fixed, whilst $\kappa $ is left as undetermined: 
\be
q(t)=\frac {r^2}{256} (t-r)^2+
\frac {3r}{256} (t -r)^3 +\kappa (t -r)^4 + O(t-r)^5 \Leftrightarrow 
\hat \eta (t) = -\ln |t-r| +\ln 2- \frac {t-r}{2r}+ \frac {56-4096\kappa}{32 r^2}(t-r)^2 + O(t-r)^3  \label {qexp}  \, .
\ee
In fact, the constant $\kappa$ and the position of the zero $r$ are the two parameters which determine the solution of the second order ODE (\ref {paintau}): the other coefficients of (\ref {qexp}) are fixed by (\ref {paintau}). 
Now, we can write the second order ODE in $y$ satisfied by $\psi _\pm (t,y)$ from the first of 
(\ref {tradlax}) (we put the phase $\phi =0$)
\ba
&& \frac{\partial ^2  \psi _{\pm}}{\partial y ^2} - \frac {1} {2} \left (\coth \left (\frac {y} {2} \pm \hat \eta \right ) \mp 1\right )\frac {\partial \psi _\pm} {\partial y}  \mp \frac {1}{8} \left (1+t\frac {d \hat \eta}{dt} \right ) \coth \left (\frac {y}{2}\pm \hat \eta \right)  \psi _{\pm} + \nonumber \\
&&+\left (\frac {1} {16}-  \frac {t^2}{16} \left ( \frac {d\hat \eta}{dt} \right )^2 \right ) \psi _{\pm}+\frac {t^2}{32}[-\cosh y +\cosh 2\hat \eta ]\psi _{\pm}=0
\label {eq-psi} \, . 
\ea
Inserting (\ref  {qexp}) into this we observe cancellations of all diverging terms, proportional to $(t-r)^{-2}$ and $(t-r)^{-1}$, which leave us with the same equation for $\psi _\pm (y)$,
the MME (\ref {math}) with\footnote {The constant $P^2$ has here a different meaning as stemming from the expansion of the Painlev\'e solution, whilst it was in (\ref {math}) associated to the parent theory, quantum SG, free energy.} $ P^2=-\frac {35}{64}+48\kappa$
\be 
\partial _y ^2 \psi _\pm (y)=\left ( \frac {r^2}{32} \cosh y +P^2 \right )\psi _\pm (y)
 \label {math2} \, ,
\ee
for
$\lim \limits _{\substack {t\rightarrow r \\ \phi \rightarrow \pi}}  \psi _\pm (y+2i\phi) =i\sqrt (8/r)^{\pm 1/2} \left (\lim \limits _{t\rightarrow r} e^{\frac {\hat \eta}{2}} \right )q^{(0)}_\pm(y+i\pi) $. 
This concludes the proof that the $TQ$-system of the P-ODE/IM above at $\xi =0$ is the MME and a particular point of the Painlev\'e III$_3$ flow. Moreover, it is the quantisation (instanton regularisation) of the square of SW differential for pure ($N_f=0$) $SU(2)$ ${\cal N}=2$ super Yang-Mills (SYM) in the NS regime (the level $2$ CFT null vector equation \cite {AY} in an AGT perspective \cite {AGT}), whose connexion and monodromy coefficients (at this particular point of the flow $t=r$) have been studied in \cite{FG} - and then for the whole Painlev\'e flow. In fact, they are time independent as the flow is {\it isomonodromic} and linked to gauge periods, with the parametrisation of the Child spectral parameter in the form $r=8e^\theta$ \cite{FG}. The latter method is named C-ODE/IM in our spirit, since it finds the connection and monodromy coefficients  of the solutions ${\cal Q}_\pm$ to the $TQ$-system derived in the P-ODE/IM; of course, it is under study here only the particular case $\xi=0$, when the $TQ$-system reduces to a differential equation, but it is very interesting for the future to study the monodromies of the full-fledged finite differential equation. In this way, the Child NS gauge theory appears as limit to the time $t=r$ of the Parent Painlev\'e theory which can be interpreted as a time-dependent (or flowing) ODE/IM correspondence and yields instead the Self-Dual (SD) theory \cite{LIS,LIS2} (cf. also next subsection (\ref {SD})). Furthermore, the crucial isomonodromy brings
the general advantage of considering the ODE arising as limit of the first $y$-Lax operator at the Painlev\'e zero/pole and studying the connexion and monodromy coefficients at this simpler point by using the decaying solutions as performed in \cite{FG}. Moreover, the latter can be linked to the Floquet solutions of (\ref{math2}), as studied in \cite {FG-BH-letter},
with Floquet index $k$ ($\epsilon =\pm 1$ takes track of the specific ODE between the two of (\ref {math2}) from which they descend)
\be
F _{\epsilon , \pm} \left (y +2i\pi \right )=e^{\pm 2i\pi k}F _{\epsilon , \pm} \left (y \right )  \, , 
\label  {floq}
\ee
which we choose normalised as $F _ {\epsilon, +}(y)=F _ {\epsilon , -}(-y)$. The asymptotic behaviour of $F _ {\epsilon , +}$ is 
\be
 \ln F_ {\epsilon , +} (y) \simeq  \frac {r}{4} e^{\pm \frac {y}{2}} \mp \frac {y}{4} \pm \frac {\varphi}{2} +c \, ,
\quad \textrm {as} \quad y \rightarrow \pm \infty \, . \label {exp2} 
\ee
Now, an important observation is that any Floquet solution produces the {\it universal} (independent of their normalisation $e^c$) phase change on the infinite interval $(-\infty, +\infty)$, similar to (\ref {floq})
\be
\pm \varphi= \lim _{y \rightarrow +\infty} [\ln F_{\epsilon , \pm} (y) -\ln F_{\epsilon , \pm}(-y)] \label {phase} \, .
\ee
In fact, the 'acquired' phase $\varphi$ can be conveniently evaluated by means of the following integral expression on $\Pi _\pm (y)= d/dy \ln F_ {\epsilon , \pm} (y)$:
\be
\pm \varphi = \int _{-\infty}^0 dy' \left ( \Pi _\pm (y')+\frac {r}{8} e^{-\frac {y'}{2}}-\frac {1}{4} \right ) +  \int _0^{+\infty} dy' \left (\Pi _\pm (y')-\frac {r}{8} e^{\frac {y'}{2}}+\frac {1}{4} \right )\,  , \label {varphidef}
\ee
which shows even better that it is a sort of generalisation of Floquet index on a non compact interval. Moreover, 
$k$ and $\varphi$ are related to the relevant quantities of the C-ODE/IM (MME) \cite {FG}. In fact, the Stokes factor or Baxter (transfer matrix) $T$ depends on the Floquet index $k$ as 
\be
T=2\cos 2\pi k \label {T} \, .
\ee
Besides, by connecting this Floquet basis (of two solutions) to the usual ODE/IM basis used in \cite{FG} we can prove\footnote{We will detail this procedure in a forthcoming work \cite {FRletter2}, but it is a simple variant of the one which has been already devised in \cite {FG-BH-letter} and \cite {FG-BH-article} for connecting the Stokes factor to the Floquet index of the ODE.} the relation (for Baxter $Q$) 
\be
Q=\frac {\sinh\varphi}{\sin 2\pi k} \label {Q} \, .
\ee
In this letter we show and analyse the approach to the NS gauge theory at the zero/pole in two cases, Painlev\'e III$_3$ and III$_1$, but we conjecture that it happens in general for any Painlev\'e system (giving rise to all the gauge theories with different flavours). 
Furthermore, as we are going to show, the two initial conditions at $t=0$ are simple functions of $k$ and $\varphi$ (given by the limiting MME), which then solve the so-called connexion problem.
The same can be derived for the two data at $t=+\infty$.

\subsection{Floquet solutions and P-ODE/IM: a solution to the connexion problem}
\label {conn-prob}

The solution (\ref {qexp}) is uniquely determined by the two initial values $l,\hat \eta _0$ characterising the $t\rightarrow 0$ behaviour: 
\be
q(t)=2^{-6} t^{2-4l}e^{-2\hat \eta _0}(1+o(1)) \quad \Rightarrow \quad \hat \eta=2l \ln t +\hat \eta _0 +o(1) \, . \label {etasmallt}
\ee
Hence, this solution establishes an implicit dependence of the zero data $(r, \kappa)$ on the initial conditions $(l, \hat \eta_0)$, which is addressed to as connexion problem. 
To make this relation more explicit it is convenient to reduce equation (\ref {eq-psi}) in normal form via the Abel transformation 
$ \phi _\pm(t,y)=\pm \frac {\sqrt {2} \left (\frac {t}{8} \right )^{\pm \frac {1}{2}}} {\sqrt {e^{\frac {y}{2}(1\mp 1)}-e^{\mp 2\hat \eta +\frac {y}{2}(-1\mp 1) }}} \psi _\pm (t,y)$:
\be
\frac{\partial ^2   \phi _{\pm}}{\partial y ^2} + \frac{5+\cosh (y \pm 2\hat \eta)}{16-16\cosh (y \pm 2\hat \eta)}  \phi _{\pm} \mp \frac {t}{8} \frac {d \hat \eta}{dt} \coth \left (\frac {y}{2}\pm \hat \eta \right)  \phi _{\pm}- \frac {t^2}{16} \left ( \frac {d\hat \eta}{dt} \right )^2  \phi _{\pm}+\frac {t^2}{32}[-\cosh y +\cosh 2\hat \eta ] \phi _{\pm}=0
\label {genbisp2} \, . 
\ee
Similarly to (\ref {eq-psi}), in the limit $t\rightarrow r$ also equations (\ref {genbisp2}) become, by virtue of (\ref  {qexp}), the MME  (\ref {math2}).
This means that Floquet solutions $F_{\epsilon, \pm} (t,y)$\footnote {Again, $\epsilon =\pm 1$ indicates the equation (\ref {genbisp2})  $F_{\epsilon, \pm} (t,y)$ solve, whereas the second index $\pm $ distinguishes the two Floquet solutions.} of (\ref {genbisp2}) when $t\rightarrow r$ flow to Floquet solutions $F _ {\epsilon , \pm}(y)$ of the MME.   
It is then natural to extend to any time $t$ the definitions (\ref {floq}, \ref {phase}) by substituting  $F _ {\epsilon , \pm}(y)$ with  $F_{\epsilon, \pm} (t,y)$, 
\ba
&&F_{\epsilon, \pm}  (t,y +2i\pi )=e^{\pm 2i\pi k} F_{\epsilon, \pm} (t,y) \label {floqt}\\
&& \pm \varphi=  \lim \limits_{y \rightarrow +\infty} [\ln F_{\epsilon, \pm} (t,y) -\ln F_{\epsilon, \pm}(t,-y)] \, , \label {phaset} 
\ea
Now, because of {\it isomonodromy}, $k,\varphi$ do not depend on time $t$ and then these parameters, specific of the limit $t=r$, can be also computed at time $t=0$.
When $t\rightarrow 0$ (\ref {genbisp2}) produce a finite result for the shifted wave functions $ \phi _\pm (t,y -2\ln t)$ and $ \phi _\pm (t,y +2\ln t)$. Indeed, the general expression satisfying the Painlev\'e equation has two free parameters, $\hat \eta =2l \ln t +\hat \eta _0 +o(1)$, and therefore (\ref {genbisp2}) reduce to well-known forms (valid for $y$ around $+\infty$ and $-\infty$, respectively)
\be
t \rightarrow 0 \ \,  , \ \ \frac{\partial ^2   \phi _{\pm}}{\partial y ^2}(t,y-2\ln t)- \left [ \frac {e^y}{64}+\left (\frac {2l\pm 1}{4} \right )^2 \right ]   \phi _{\pm}(t,y-2\ln t)=0 \quad (y>-a \, , \ \ a>0) \label {kink1}
\ee
\be
t \rightarrow 0 \ \,  , \ \ \frac{\partial ^2   \phi _{\pm}}{\partial y ^2}(t,y+2\ln t)- \left [ \frac {e^{-y}}{64}+\left (\frac {2l\mp 1}{4} \right )^2 \right ]   \phi _{\pm}(t,y+2\ln t)=0 \quad (y<a) \, , \label {kink2}
\ee
with (Floquet) solutions  $ F _{\pm , \mp}(t,y-2\ln t)=J_{ -\left (\frac {1}{2}\pm l \right )}\left (\frac {i}{4}e^{\frac {y}{2}} \right ) $ (of (\ref {kink1}) and $ F_{\pm, \mp} (t, y+2\ln t)=J_{ -\left (\frac {1}{2}\mp l \right )}\left (\frac {i}{4}e^{-\frac {y}{2}} \right ) $ (of (\ref {kink2}) or $ F _{\pm , \pm}(t,y-2\ln t)=J_{ \left (\frac {1}{2}\pm l \right )}\left (\frac {i}{4}e^{\frac {y}{2}} \right ) $ (of (\ref {kink1})  and  $ F_{\pm, \pm} (t, y+2\ln t)=J_{ \left (\frac {1}{2}\mp l \right )}\left (\frac {i}{4}e^{-\frac {y}{2}} \right ) $ (of (\ref {kink2}). However, since $-\frac {1}{2}<l<\frac {1}{2}$, in the limit $t\rightarrow 0$ $F_{\pm ,\pm}(t,y)$ is subleading with respect to $F_{\pm ,\mp}(t,y)$: then, we consider the first choice, $ F_{\pm , \mp}$. A shift of $2\pi i$ produces 
\be
F_{\pm,\mp} (t,y-2\ln t+2\pi i)=e^{\left (\mp i\pi l -\frac {i\pi}{2}\right )}   F_{\pm,\mp} (t,y-2\ln t) \, , \quad 
F_{\pm,\mp} (t,y+2\ln t+2\pi i)=e^{\left (\mp i\pi l +\frac {i\pi}{2}\right )}   F_{\pm,\mp} (t,y+2\ln t) 
\ee
from which the value of the Floquet index (\ref {floqt}) with $\epsilon =\mp$ 
\be
k= \frac {l}{2}+\frac {1}{4} \, , \label {klrel}
\ee
for all $t$. 

Let us compute the acquired phase for
$F_{\pm, \mp}$: we first compute the quantity
\be
\varphi _{\pm, \mp}  = \int _{-\infty}^0 dy \left (\Pi _{\pm, \mp} (t,y)+\frac {t}{8}e^{-\frac {y}{2}}-\frac {1}{4} \right ) +
 \int _0^{+\infty} dy \left (\Pi _{\pm ,\mp} (t,y)-\frac {t}{8}e^{\frac {y}{2}}+\frac {1}{4} \right )  \label {tildephipm} \, .
\ee
where $\Pi _{\pm, \mp} (t,y)=\frac {d}{dy} \ln F _{\pm, \mp} (t,y)$. In fact, 
\be
\varphi _{\pm, \mp}= \int _{-\infty}^{-2\ln t} dy \left (\Pi _{\pm , \mp} (t,y+2\ln t)+\frac {1}{8}e^{-\frac {y}{2}}-\frac {1}{4} \right ) +
 \int _{2\ln t}^{+\infty} dy \left (\Pi _{\pm ,\mp} (t,y-2\ln t)-\frac {1}{8}e^{\frac {y}{2}}+\frac {1}{4} \right ) \, , \nonumber
\ee
is computable at $t\rightarrow 0$ from the Riccati solutions at leading order,  $\Pi _{\pm ,\mp} (t,y-2\ln t)=\frac {d}{dy}\ln J_{-\left (\frac {1}{2}\pm l \right )}\left (\frac {i}{4}e^{\frac {y}{2}} \right )$,
$\Pi _{\pm ,\mp} (t,y+2\ln t)=\frac {d}{dy}\ln J_ {-\left (\frac {1}{2}\mp l \right )}\left (\frac {i}{4}e^{-\frac {y}{2}} \right )$, which give around $t=0$
\be
\varphi _{\pm, \mp}= \pm 2l \ln t \mp \ln \frac {\Gamma \left ( \frac {1}{2}+ l \right )}{\Gamma \left (\frac {1}{2}- l \right )}\mp 2l \ln 8 +o(1) \, .
\ee


On the other hand the vector $\Psi (t,y)$ is also subjected to the equation in $t$ $(\partial _t - {\bf A}_t)\Psi  (t,y) = 0$. On $\phi _\pm$ this implies
\be
\left [ \frac{\partial ^2  }{\partial t ^2} \mp \tanh \left (\frac {y}{2} \pm \hat \eta \right ) \frac {d\hat \eta}{dt} \frac {\partial }{\partial t}-\frac {1}{8}(\cosh y +\cosh 2\hat \eta ) \right ]  \sqrt {\sinh \left (\frac {y}{2}\pm \hat \eta \right )}\phi _\pm (t,y)=0 \label {tiequation}
\ee
The study of both (\ref {genbisp2}) and (\ref {tiequation}) in the limits $ y\rightarrow \pm \infty$ produces on Floquet solutions the constraints 
\be
y\rightarrow +\infty \Rightarrow  F_{\epsilon ,\mp}(t,y) \simeq e^{\mp \frac {\tilde \varphi _{\epsilon}}{2}}e^{-\epsilon \frac {\hat \eta}{2}} e^{\frac {t}{4}e^{\frac {y}{2}}-\frac {y}{4}}  \  ; \quad y\rightarrow - \infty \Rightarrow  
F _{\epsilon ,\mp} (t,y) \simeq e^{\pm \frac {\tilde \varphi _{\epsilon}}{2}} e^{\epsilon \frac {\hat \eta}{2}} e^{\frac {t}{4}e^{-\frac {y}{2}}+\frac {y}{4}  }
\ee 
with the quantity $\tilde \varphi _{\epsilon}$ independent of neither $t$ nor $y$. 
Then, in order to get the acquired phase for $F_{\epsilon ,\mp}$ on should add $-\epsilon \hat \eta $ to $\varphi _{\epsilon ,\mp}$. Then, for our Floquet solutions $F_{\pm, \mp}$ we find that the acquired phase 
\be
\mp \varphi = \varphi _{\pm, \mp} \mp \hat \eta = \mp \hat \eta _0 \mp \ln \frac {\Gamma \left ( \frac {1}{2}+ l \right )}{\Gamma \left (\frac {1}{2}- l \right )}\mp 2l \ln 8  \, , 
\ee
does not depend on $t$. 
Now, the initial constants are expressed in terms of the data at the zero/pole: 
\be
l=2k-\frac {1}{2} \, , \quad \hat \eta _0=\varphi -\ln \frac {\Gamma (2k)}{\Gamma (1-2k)}- (4k-1)\ln 8 \quad \Rightarrow \quad 
e^{2\hat \eta}=e^{2\varphi} \frac {\Gamma ^2(1-2k)}{\Gamma ^2(2k)} \left (\frac {t}{8} \right )^{8k-2}[1+o(1)] 
\label {e2eta} \, .
\ee
This result eventually proves that the indexes $k$ and $\varphi$ of the Floquet solutions have a crucial r\^ole in the parent theory as they fix the two initial (and final) conditions at $t=0$ (and $t=+\infty$, respectively) of the Painlev\'e equation. Given these, the rest of the expansion, at small (large) $t$ can be computed recursively by using the differential equation itself. In fact, upon writing $q(t)=2^{-6}t^{2-4l}e^{\beta (t)}$, it enjoys the form
\be
\beta (t)=\sum _{n,m=0}^{+\infty}B_{n,m}t^{\alpha _{n,m}} \, , 
\ee
with $\alpha _{n,m}=(2+4l)n+(2-4l)m$; moreover, $B_{0,0}=-2\hat \eta _0$ and the others $B_{n,m}$ follow from the relation
\be
2\sum _{n,m=0}^{+\infty}B_{n,m}\alpha _{n,m}^2t^{\alpha _{n,m}-2}=t^{-4l}\exp \left (\sum _{n,m=0}^{+\infty}B_{n,m}t^{\alpha _{n,m}}\right )-t^{4l}\exp \left (-\sum _{n,m=0}^{+\infty}B_{n,m}t^{\alpha _{n,m}} \right ) \, . \label {recB}
\ee
Now, the connection problem can be solved exactly, as we can express $k,\varphi$, i.e. $l,\hat \eta _0$ through (\ref {e2eta}) (or similarly for the two data at $t=+\infty$) in terms of the parameters $r=8e^\theta $ and $P^2$ of the MME (\ref{math2}), ruling the Painlev\'e zero/pole (\ref{qexp}). In fact, we know that in the child theory the transfer matrix (Stokes coefficient) can be easily given by 
$T=2\cos 2\pi k=Q(\theta -i\pi/2,P^2)Q(\theta +i\pi,P^2)-Q(\theta +i\pi/2,P^2)Q(\theta,P^2)$
and moreover the $TQ$-system takes the form $2 \cot 2\pi k \sinh \varphi=Q(\theta +i\pi/2,P^2)+ Q(\theta-i\pi/2,P^2)$ \cite{FG}; eventually $Q(\theta,P^2)$ is given by the integrability (or gauge) TBA \cite{FG}. Alternatively, we can use from \cite{FG-BH-letter, FG-BH-article} that $k,\varphi$ do coincide respectively with the gauge period $a=k(r,P^2)$ computed as Floquet index\footnote {Alternatively, we can use the Matone relation $P^2=-\frac {r}{2} \frac {\partial {\cal F}^{NS}}{\partial r}$.}  and its dual\footnote {This identification will be proven in  \cite {FRletter2}, together with an analogous one for the $N_f=2$ case (Painlev\'e III$_1$ below).\label {foot}}
\be
A_D=\partial {\cal F}^{NS}/\partial a=\varphi \label {adphi}
\ee
({\it cf.} the expression (\ref {Q}) for the connection coefficient), thus given by the gauge prepotential ${\cal F}^{NS}(r,a)$. Therefore, the latter  gives an explicit solution to the connexion problem, as well. Furthermore, 
we will present in a forthcoming publication \cite {FRletter2} another exact way to compute $k$ and $\varphi$ from the MME (\ref{math2})
within integrability.

\subsection {From Child (NS) to Parent (SD) gauge theory}
\label {SD}

As the more general classical ShG \cite {FR1}, the Painlev\'e III$_3$ parent theory can be described in terms of the tau-function $\tau (s)$ \cite {LUK}, giving $q(s)$ via $\frac {s}{q(s)} =-\left (s \frac {d}{ds} s \frac {d}{ds} \right ) \ln \tau (s)$. From recursion relations (\ref {recB}), concerning $q(t)$, one can argue those for the coefficients of the tau-function at small times:
more deeply and interestingly, the tau-function expression (4.3) of \cite {FR1} should yield a good limit function in $\xi =0$, valid for the Painlev\'e III$_3$ equation. Unfortunately, 
we have not found this limit yet, therefore we have to borrow an expression from the landmark paper \cite {LIS}\footnote {We use notations of \cite {LIS2}: comparing (3.1) of \cite {LIS2} with (\ref {e2eta}), we find that $\sigma$ and $\eta$ of that paper equal $k$ and $\varphi/2\pi i -1/2$, respectively. (\ref {taubis}) is a rewriting of (3.3) in \cite {LIS2} in terms of $k$ and $ \varphi$.}
\be
\tau (s)=\sum _{n=-\infty}^{+\infty}  \exp \left [2n \varphi + {\cal F}^{SD}(k + n, s) \right ] \, , \label {taubis}
\ee
where $\exp \left ({\cal F}^{SD} (k, s) \right)$ is the Nekrasov partition function in the SD $\Omega$-background $\epsilon_1=-\epsilon_2$ (or the irregular $c=1$ conformal block), with Coulomb branch v.e.v. $k$ and instanton coupling $s$ (corresponding to the time), whereas for us here it is an unknown function to be fixed by the recursion relation for the coefficients of the small time $s$ expansion. A similar approach has been used in \cite {BGT}, but here we want to add that the simplifying shifted argument $k+n$ in (\ref {taubis}) \cite {LIS} can be deduced from the regularity pattern of the recursion, like in $\alpha _{n,m}$ of (\ref {recB}). Finally, upon inserting into (\ref {taubis}) the
double zero of $q(t)$ 
at $t=r$, which means $\tau (s=(r/8)^4)=0$ (with $r=r(k,\varphi)$ derived above), we can derive the celebrated blowup equation \cite {NY} in a novel way:
\be
\sum _{n=-\infty}^{+\infty} \exp \left [2n \varphi +  {\cal F}^{SD} \left (k + n, \frac {r^4}{8^4} \right) \right ]=0 \, ,
\label {blow-up}
\ee 
with $\varphi =\partial { {\cal F}^{NS}}/\partial a=A_D$, as stated in (\ref {adphi}).


\section{The Painlev\'e III$_1$ theory}
\label{P31}
\setcounter{equation}{0}

We need now to repeat our method for the Painlev\'e III$_1$, with particular attention to what happens when the Painlev\'e flow $t$ reaches a singularity.
The Painlev\'e III$_1$ equation arises as isomonodromic deformation from the compatibility condition of two first order differential Lax operators $[\partial _z-{\bf A}_z,\partial _s - {\bf A}_s]=0$, between the first 
Lax linear problem $\partial _z \psi ={\bf A}_z \psi $, where ${\bf A}_z= \begin{pmatrix}  A&B \\C&D \end{pmatrix}$, with $z$ auxiliary parameter and 
\be
A=-D=\frac {\sqrt {s}}{2}-\frac {\theta _{\ast}}{z}+\frac {\sqrt {s}(2p-1)}{2z^2} \, , \quad B=-\frac {pqu}{z}-\frac {\sqrt {s}pu}{z^2} \, , \quad C=\frac {pq-p^2q+\theta _{\ast}+\theta _{\star}-2p\theta _{\ast}}{puz}+\frac {\sqrt {s}(p-1)}{uz^2} \label {abcd}
\ee
and the time linear Lax problem $\partial _s \psi ={\bf A}_s \psi $, where ${\bf A}_s= \begin{pmatrix}  \tilde A& \tilde B \\\tilde C&\tilde D \end{pmatrix}$, with $s$ time and 
\be
\tilde A=-\tilde D=\frac {z}{4\sqrt {s}}+\frac {1-2p}{4\sqrt {s}z} \, , \quad \tilde B=-\frac {pqu}{2s}+\frac {pu}{2\sqrt {s}z} \, , \quad \tilde C=\frac {pq-p^2q+\theta _{\ast}+\theta _{\star}-2p\theta _{\ast}}{2spu}+\frac {1-p}{2\sqrt {s} uz} \label {tilde abcd} \, .
\ee
It writes explicitly for the unknown functions $q(s),p(s),u(s)$ as
\be
s \dot u=u\left (\theta _{\ast}-\frac {\theta _{\ast}+\theta _{\star}}{p}-q \right ) \, , \quad s\dot q=s-{q^2}+{2pq^2}+{2q\theta _{\ast}} \, \quad s \dot p=2pq-2p^2q+\theta _{\ast}+\theta _{\star}-2p\theta _{\ast} \, .
\ee
In fact, they lead to the Painlev\'e III$_1$ equation for $q(s)$
\be 
\ddot q=\frac {\dot q ^2}{q}-\frac {\dot q}{s}+\frac {q^3}{s^2}+\frac {2q^2\theta _{\star}}{s^2}+\frac {1-2\theta _{\ast}}{s}-\frac {1}{q} \, , 
\ee
which has two parameters and the symmetry $q(s,\theta _{\star},\theta _{\ast}) \rightarrow s/q(s,\theta _{\ast}-1/2,\theta _{\star}+1/2)$. Even though in the case of Painlev\'e III$_1$ we still do not have a {\it direct} understanding of the P-ODE/IM correspondence, we can study the first order Lax ODEs, which for the two components of the vector $\psi = \begin{pmatrix}  \psi _1 \\ \psi _2  \end{pmatrix}$ write in the (separated) second order form
\ba
\partial ^2_z \psi _1&=&(\partial _z A+BC-A \frac {\partial _z B}{B}+A^2)\psi _1 + \frac {\partial _z B}{B}\partial _z \psi _1 \, ,
\label {eqmasterter} \\
\partial ^2_z \psi _2&=&(-\partial _z A+BC+A \frac {\partial _z C}{C}+A^2)\psi _2 + \frac {\partial _z C}{C}\partial _z \psi _2 \, ,
\label {eqmasterter2}
\ea
with attention on their connection and monodromy coefficients. With this aim it is important to see what happens to them at a peculiar time, namely when $q$ reaches a zero or pole. Since it is not clear how the most suitable time $t$ is linked to $s$, it can be defined in general as $t =\xi s^{\alpha}$, with $\xi, \alpha$ real. Then the Painlev\'e III$_1$ has the form
\be 
\alpha ^2 t^2 q^{''}+\alpha ^2 t q^{'}=\alpha ^2 t ^2 \frac {{q^{'}}^2}{q}+q^3+2q^2 \theta_{\star}+(1-2\theta _{\ast})t ^{\frac {1}{\alpha}}\xi ^{-\frac {1}{\alpha}}
-\frac {1}{q} t^{\frac {2}{\alpha}}\xi ^{-\frac {2}{\alpha}}
\label  {P31eq} \, , 
\ee
where $q'=dq/dt$. 
If a point $t=r$ is a pole or equivalently, because of the above symmetry, a zero of the Painlev\'e solution $q$, equation (\ref  {P31eq}) constraints it to be simple. 
The expansion of $q(t)$ as a solution of (\ref {P31eq}) around $t=r$ is
\be 
q(t)=\frac {\alpha r}{t-r}+\frac {\alpha -2\theta _{\star}}{2}+\zeta (t-r)+O\left ((t -r)^2\right ) \, ,  
\label {q31}
\ee
where $\zeta$ is an arbitrary constant. As a consequence of this, $2pq=-2(\theta _{\ast}+\theta _{\star})+\omega (t -r)+ O((t-r)^2) $, with $\omega =3\zeta-\frac {r^2(\alpha -2\theta _{\star})^2}{4\alpha r^3}+\frac {\alpha -2\theta _{\star}}{2r}-
\frac {r^{\frac {1}{\alpha}}\xi ^{-\frac {1}{\alpha}}}{\alpha r}$ and then the Lax entries expand as  
\ba
&& \frac {\partial _z B}{B}=-\frac {1}{z}+O(t-r) \, , \quad \partial _z A=\frac 
{\theta _{\ast}}{z^2}+\frac {r^{\frac {1}{2\alpha}}\xi ^{-\frac {1}{2\alpha}}}{z^3}+O(t-r) \, , \quad  \frac {\partial _z C}{C}=-\frac {1}{z}+O(t-r) \, , \nonumber \\
&& A=\frac {r^{\frac {1}{2\alpha}}\xi ^{-\frac {1}{2\alpha}}}{2}-\frac {\theta _{\ast}}{z}-\frac {r^{\frac {1}{2\alpha}}\xi ^{-\frac {1}{2\alpha}}}{2z^2}+O(t-r) \, , \quad BC=-\frac {r^{\frac {1}{2\alpha}}\xi ^{-\frac {1}{2\alpha}}}{z^3}(\theta _{\ast}+\theta _{\star})-\frac {1}{2z^2}\left (\omega \alpha r +2(\theta _{\ast}^2-\theta _{\star}^2) \right )+O(t-r) \nonumber \, . 
\ea
Eventually, in the limit $t\rightarrow r$ the two components $\psi_1, \psi_2$ satisfy the simplified ODEs
\ba
\partial _z^2 \psi _1 +\frac {1}{z}\partial _z \psi _1 &=&\left ( \frac {2e^\theta q_1}{z}+\frac {P^2}{z^2}+\frac {2e^\theta q_2}{z^3}+\frac {e^{2\theta}}{z^4}+e^{2\theta}\right )\psi _1 \, ,  \\
\partial _z^2 \psi _2 +\frac {1}{z}\partial _z \psi _2 &=& \left ( \frac {2e^\theta (q_1-1)}{z}+\frac {P^2}{z^2}+\frac {2e^\theta (q_2-1)}{z^3}+\frac {e^{2\theta}}{z^4}+e^{2\theta} \right )\psi _2 \, , 
\ea
where
\be
2e^\theta q_1=r^{\frac {1}{2\alpha}}\xi ^{-\frac {1}{2\alpha}}\left (\frac {1}{2}-\theta _{\ast}\right ) \, , \quad 
P^2=2\theta _{\ast}^2-\frac  {1} {2}\theta _ {\star}^2-\frac  {1}  {8}\alpha ^2-\frac  {3} {2}\alpha \zeta r \, , \quad
2e^\theta q_2=r^{\frac {1}{2\alpha}}\xi ^{-\frac {1}{2\alpha}}\left (\frac {1}{2}-\theta _{\star}\right ) \, ,\quad
e^{2\theta}=\frac {r^{\frac {1}{\alpha}}\xi ^{-\frac {1}{\alpha}}}{4}  
 \, . 
\ee
Upon changing variable $z=e^y$, they both acquire the form of the DCHE
\ba 
\partial ^2 _y \psi _1&=&\left [ e^{2\theta}(e^{2y}+e^{-2y})+2e^\theta q_1 e^y+2e^\theta q_2 e^{-y}+P^2 \right ] \psi _1 \, , 
\label {firstcomp} \\
\partial ^2 _y \psi _2&=&\left [ e^{2\theta}(e^{2y}+e^{-2y})+2e^\theta (q_1-1)e^y+2e^\theta (q_2-1)e^{-y}+P^2 \right ] \psi _2 \, ,
\label {second comp}
\ea
which is the NS quantisation of the SW differential for ${\cal N}=2$ SYM with $N_f=2$ flavours \cite {MT}.
Now, we can repeat what we did for the MME/$N_f=0$ case: after defining the Floquet solutions of (\ref {firstcomp}) (similar definitions can be done for (\ref {second comp})), we can identify their Floquet index $k$ and the 'acquired' phase (symmetrised w.r.t. the two masses $q_1,q_2$) $\varphi$ as, respectively, the gauge period $a$ and its dual $A_D$ of ${\cal N}=2$ SYM with $N_f=2$ in NS background, link them to the {\it isomonodromic} connexion coefficient $Q$ and the latter to the two initial data. This construction will be reported in a subsequent publication \cite {FRletter2}.

\section{Summary and perspectives}
\label{concl}

We have presented and analysed two examples of a novel simple occurrence of Painlev\'e theory inside NS regime ${\cal N}=2$ supersymmetric $SU(2)$ gauge theories. For what concerns Painlev\'e III$_3$ theory, its first differential Lax operator in $z$, first of (\ref {tradlax}), reduces at the Painlev\'e singularity time $t=r$ to the MME, (\ref {math2}). It is not a coincidence that the latter is the quantum (NS regime) SW differential for ${\cal N}=2$ supersymmetric gauge theories with no flavour ($N_f=0$).
In fact, the same limit is performed for the Painlev\'e III$_1$: the ODE deriving from the first Lax operator in $z$ (\ref {eqmasterter}, \ref {eqmasterter2}) as limit to the Painlev\'e singularity $t=r$ is the DCHE (\ref {firstcomp}, \ref {second comp}), corresponding to the quantum SW differential for ${\cal N}=2$ supersymmetric gauge theories with $N_f=2$ flavours.
In general, we conjecture that this limit gives, on each Painlev\'e equation III, V, VI, the gauge theories with different $N_f=0,1,2,3,4$, respectively, in NS background. Other Painlev\'e equations (I, II and IV) give rise to Argyres-Douglas theories.

An important consequence of this limit comes from the isomonodromy of the Painlev\'e flow. In fact, these quantum SW differentials have the same monodromies of the first (in variable $z$) Lax operators  of Painlev\'e equations. Explicitly the monodromy and connexion coefficients $T$ and $Q$, respectively, are time independent and have been already studied in the C-ODE-IM correspondences of \cite {FG,FG-BH-letter,FG-BH-article} by connecting them to the SW periods. In this letter we added to this set-up the explicit expression (\ref {varphidef}), the acquired phase $\varphi$\footnote {This can be defined similarly for the Floquet solutions of (\ref {firstcomp}).}, for the dual instanton period. It can be studied in all possible regimes, not only for small instanton coupling. In particular it can be linked to the isomonodromic connexion coefficient $Q$ (as shown for instance in (\ref {Q}). In fact these relations show that in general it is not time invariant, except in the case $N_f=0$. Nevertheless the computation of the same $Q$ at time $t=0$ (or $t=+\infty$) gives a solution to the connexion problem. 

In a reverse perspective, we can say that isomonodromic deformations of quantum SW differentials give Lax pairs for Painlev\'e equations; a similar approach was pursued in \cite {CLT}.
In this context the correspondence of the Painlev\'e solutions to the SD gauge partition functions
\cite {LIS,LIS2} needs further understanding, beyond the new way to the blow-up equations found above.   
Besides, the immediate continuation of this work will also concern a deeper C-ODE/IM correspondence for the MME and DCHE, with the idea of computing explicitly the connexion and monodromy coefficients $Q$ and $T$, the acquired phase $\varphi $, or in other words the dual instanton period $A_D$, and finally the gauge prepotential  ${\cal F}^{NS}$. The latter would give explicit solution to the connexion problem and to the ODE/IM correspondence.

\medskip

{\bf Acknowledgements} 
Discussions with G. Bonelli, D. Gregori, P. Longhi, A. Tanzini are kindly acknowledged.  
This research was supported in part by the grants GAST (INFN), the MPNS-COST Action MP1210, the EC Network Gatis, the MIUR-PRIN contract 2017CC72MK\textunderscore 003 and the National Science Foundation under Grant No. NSF PHY-1748958.

\end{document}